\documentclass[prl,twocolumn,superscriptaddress]{revtex4}
\usepackage{color}
\usepackage{graphicx}
\usepackage{graphics}
\begin{document}

\title{ Structural fuzziness of large gold clusters}

\author{Kuo Bao}
\affiliation{ State Key Lab of Superhard Materials, Jilin University, Changchun 130012, China \\}
\affiliation{Departement of Physics, Universit\"{a}t Basel, Klingelbergstr. 82, 4056 Basel, Switzerland \\}
\author{Stefan Goedecker}
\affiliation{Departement of Physics, Universit\"{a}t Basel, Klingelbergstr. 82, 4056 Basel, Switzerland \\}
\author{Kenji Koga}
\affiliation{ National Institute of Advanced Industrial Science and Technology (AIST),
Tsukuba, Ibaraki 305-8565, Japan}
\author{ Fr\'{e}d\'{e}ric Lan\c{c}on }
\affiliation{CEA, INAC, SP2M, Laboratoire de simulation atomistique (L\_Sim), F-38054
Grenoble, France}
\author{Alexey Neelov}
\affiliation{Departement of Physics, Universit\"{a}t Basel, Klingelbergstr. 82, 4056 Basel, Switzerland \\}

\begin{abstract}
The energetic ground state of gold clusters with up to 314 atoms
consists of rather complicated geometries that have only a weak
resemblance to the perfect icosahedra, decahedra and octahedra that are
encountered for some magic numbers. The
structure changes in most cases completely by the addition of a
single atom. Other low energy structures are so close in energy that
their Boltzmann weight is not negligible at room temperature.

\end{abstract}

\pacs{PACS numbers: ?? }

\maketitle

Since the structure determines the functionality of any material,
the knowledge of the structure is the first step to understand any
property of a condensed matter system.  Clusters are a particularly
interesting system both scientifically and technologically. It is
well known that for small clusters of less than 30 atoms the
structure can drastically change by the addition of a single
atom~\cite{landman, zeng,hellmann1}. Size selected clusters offer thus the
possibility to obtain a large variety of structures with possibly
widely different functionalities. It has for instance been revealed that
the reactive properties of Au$_{55}$ is drastically different from
other gold clusters~\cite{Au55}.

By applying global optimization
methods to find the geometric ground state structure it has also been
shown that the structures of small clusters are in most cases
amorphous~\cite{garzon,zeng,hellmann1}.
Global geometry optimization have been performed in a systematic way
for up to 80 atoms~\cite{doye} and for selected clusters up to
75 atoms~\cite{garzon}. For such a small number of atoms most
clusters are amorphous. i.e they have no well defined structure.
This feature is independent of the exact form of the potential
which is the Rosato Guillop\'{e} Legrand potential~\cite{rgl} (RGL)
in our case, the Sutton-Chen
potential in reference~\cite{doye} or the Gupta potential in
reference~\cite{garzon}. We extend these studies to much larger cluster
sizes for which we no longer find amorphous ground state geometries.
In order to determine the structure of large clusters containing a few hundred atoms,
educated guesses of low energy structures were until now made instead of systematic
global optimization. This procedure was used
for metallic clusters in many publications. Both for model
Lennard-Jones clusters~\cite{raoult}
as well as for gold clusters described by various potentials~\cite{landman,baletto,carvalho}
they arrived all at
the same conclusion. For small clusters icosahedral structures are
energetically the lowest, for medium size clusters decahedral
structures are best and for large clusters finally truncated octahedra. The
reason for this evolution of the shape is that the ratio of surface
to volume atoms is  decreasing as the cluster size increases.
Icosahedra are terminated everywhere with the energetically most
favorable (111) surfaces and they are nearly spherical, but the
internal atoms have a lot of strain. The truncated octahedra on the other
hand have inside the perfect fcc crystalline structure but
(100) facets with large surface energy. The decahedra
have an intermediate behavior. The cluster sizes at which the
transition between the different structures takes place depends on
the element~\cite{review}. For gold the transitions happen already
at rather small cluster sizes~\cite{baletto}.

\begin{figure}[h]             
\begin{center}
\setlength{\unitlength}{1cm}
\begin{picture}( 11.,6.)           
\includegraphics[width=8cm]{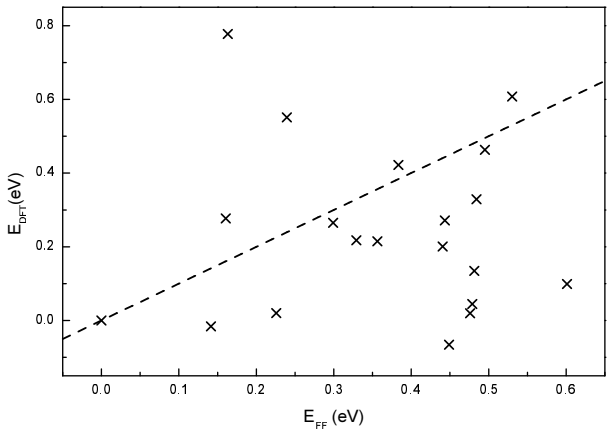}
\end{picture}
\caption{ \label{RGLDFT} Comparison of the density functional (DFT) and RGL force field (FF)
energies for relaxed configurations. The geometry relaxation with the RGL potential
did only slightly change the stable DFT structures.
The DFT configurations were obtained by doing
Minima hopping with a density functional program~\cite{bigdft}. If all the data
points were on the dashed line we would have perfect agreement between DFT and FF
for the energy differences between different structures. }
\end{center}
\end{figure}

In this study we have obtained the structure of gold clusters with
up to 300 atoms by global optimization using the minima hopping
method~\cite{minhop,roy}. Minima hopping is a highly efficient global optimization
method which can in particular reliably find the global minimum for
multi-funnel landscapes. Since global optimization on the density
functional level is not possible for such cluster sizes we have used
the RGL interatomic potential with the
parameters of reference~\cite{baletto}, which has been widely used
in studies of metallic clusters and turned out to be rather
reliable. Fig.~\ref{RGLDFT} shows the accuracy compared to density
functional calculations for several configurations of a Au$_{20}$
cluster. The errors are of the order of 1 eV which is a typical
value for high quality force fields or tight binding schemes~\cite{hellmann1}.

\begin{figure}[h]             
\begin{center}
\setlength{\unitlength}{1cm}
\begin{picture}( 11.,6.)           
\hspace{-0.2cm} \vspace{-4cm} \includegraphics[width=9cm]{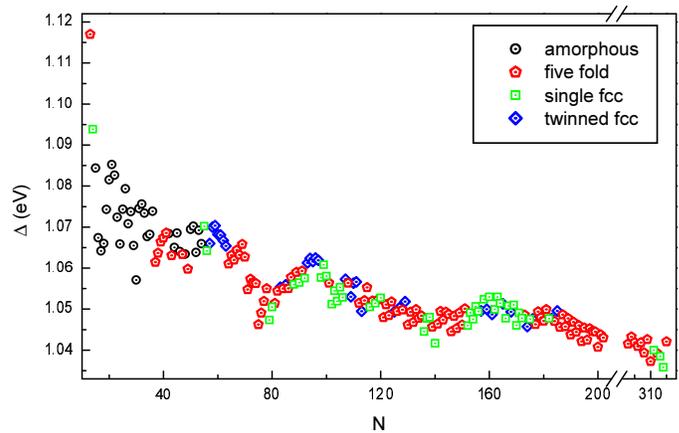}   
\end{picture}
\caption{ (colored online) \label{energies} The quantity $\Delta$ of Eq.~\ref{delta}
for all the clusters that were studied. Among all the 5-fold clusters only Au$_{13}$ is
an icosahedron. }
\end{center}
\end{figure}

\begin{figure}[h]             
\begin{center}
\setlength{\unitlength}{1cm}
\begin{picture}( 11.,10.8)           
\put(0.0,8.0){\includegraphics[width=2.54cm]{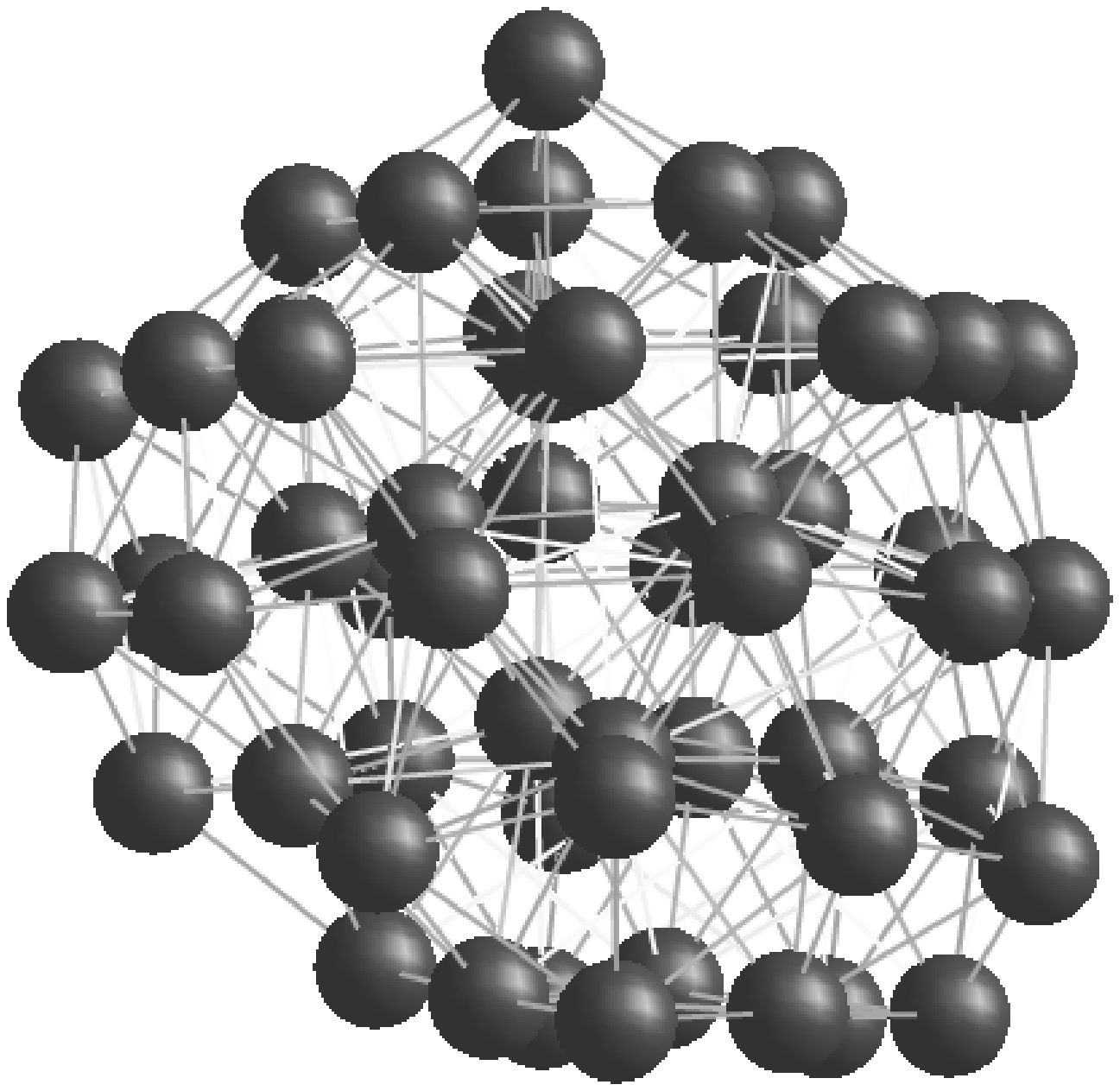}}
\put(2.58,8.0){\includegraphics[width=2.54cm]{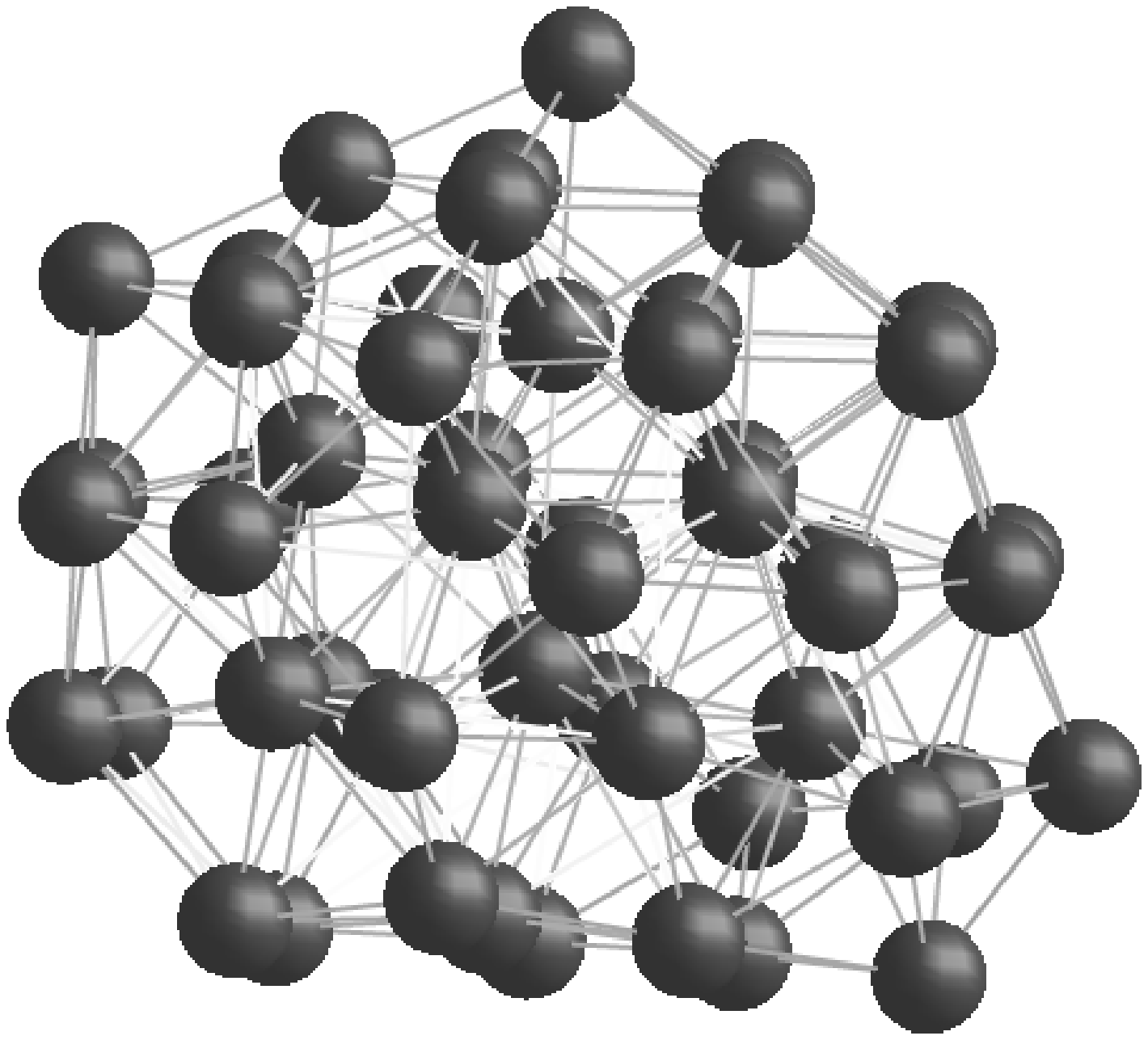}}
\put(5.12,8.0){\includegraphics[width=2.54cm]{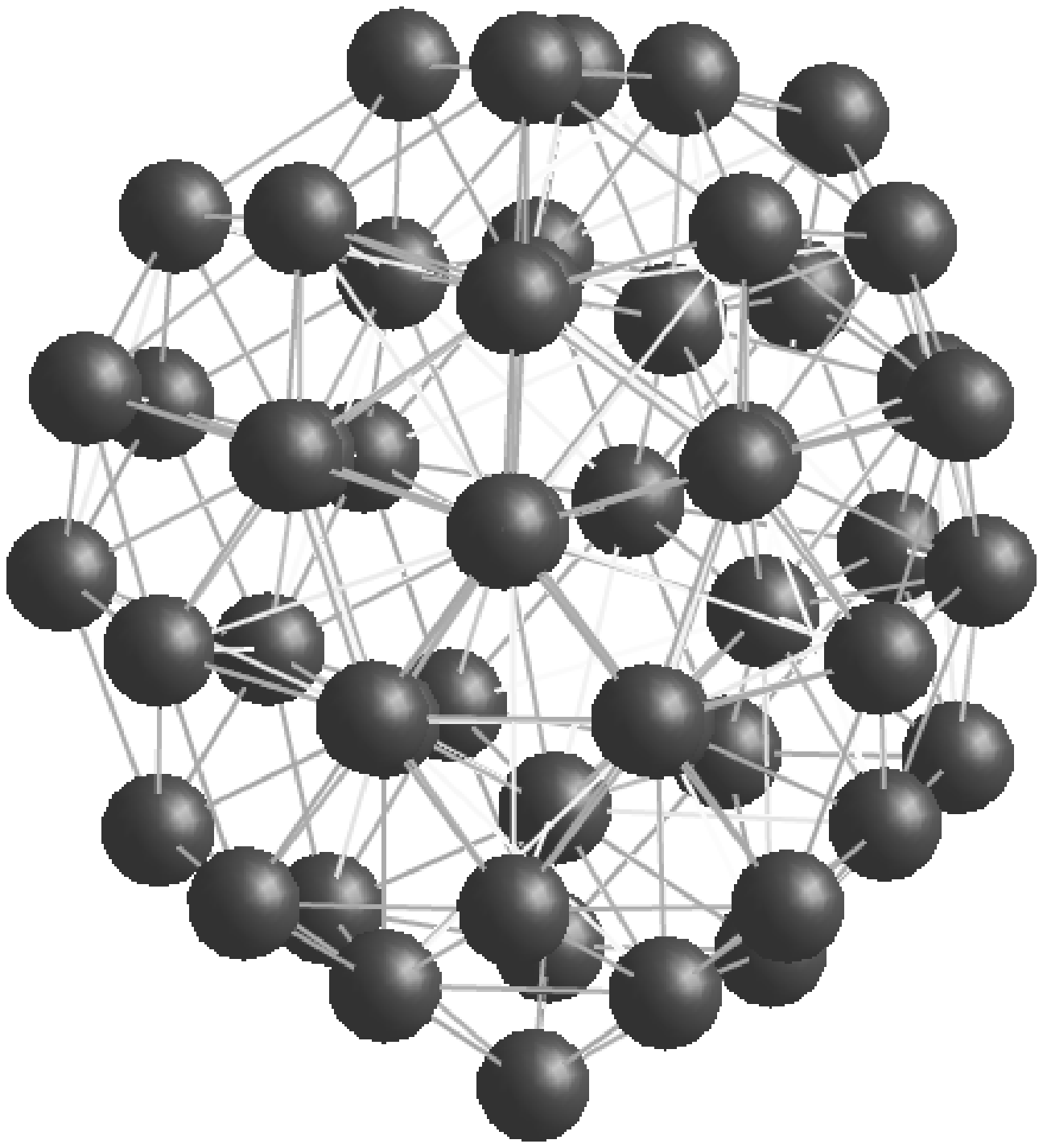} }  

\put(0.0,5.3){\includegraphics[width=2.54cm]{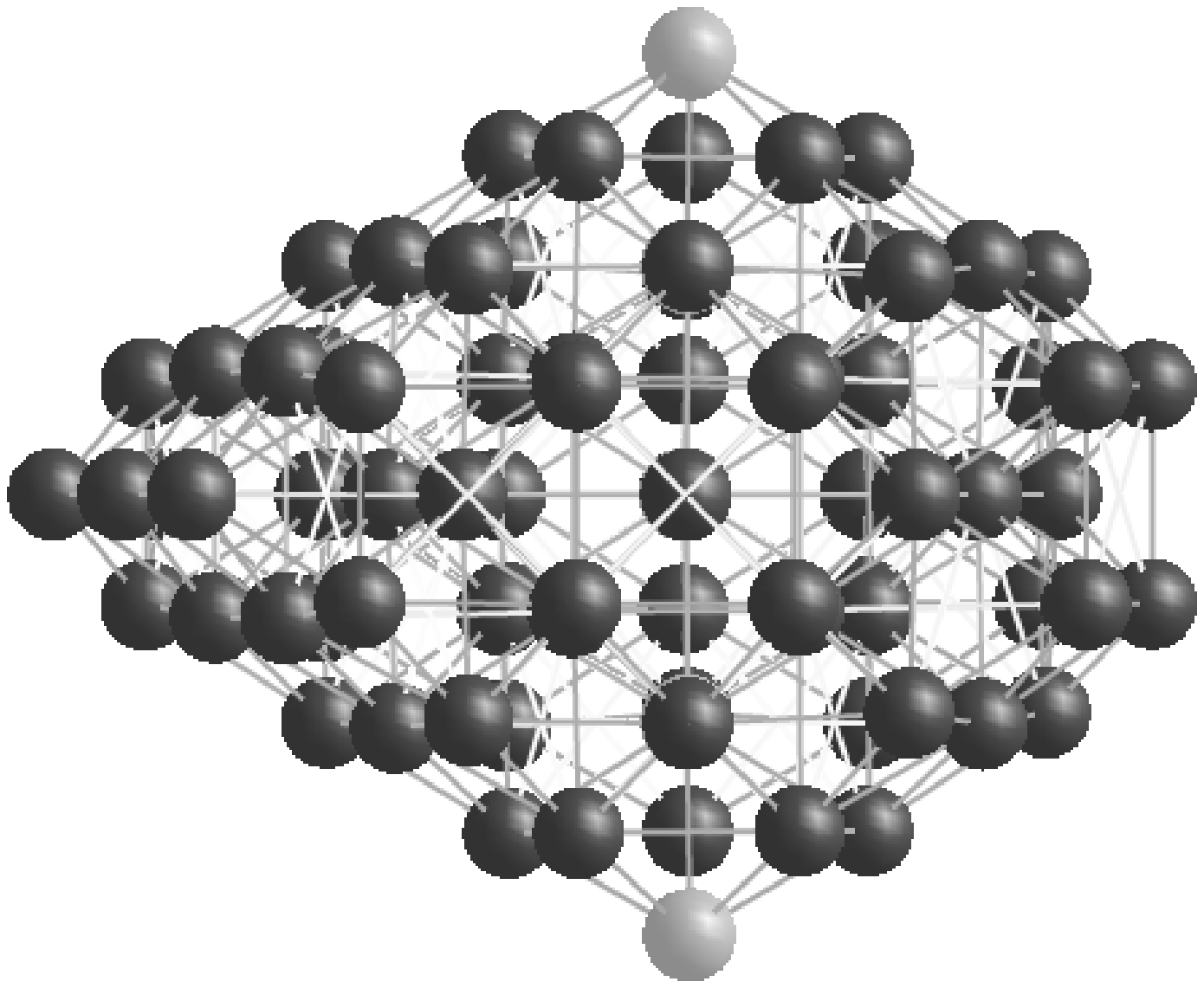}}
\put(2.58,5.3){\includegraphics[width=2.54cm]{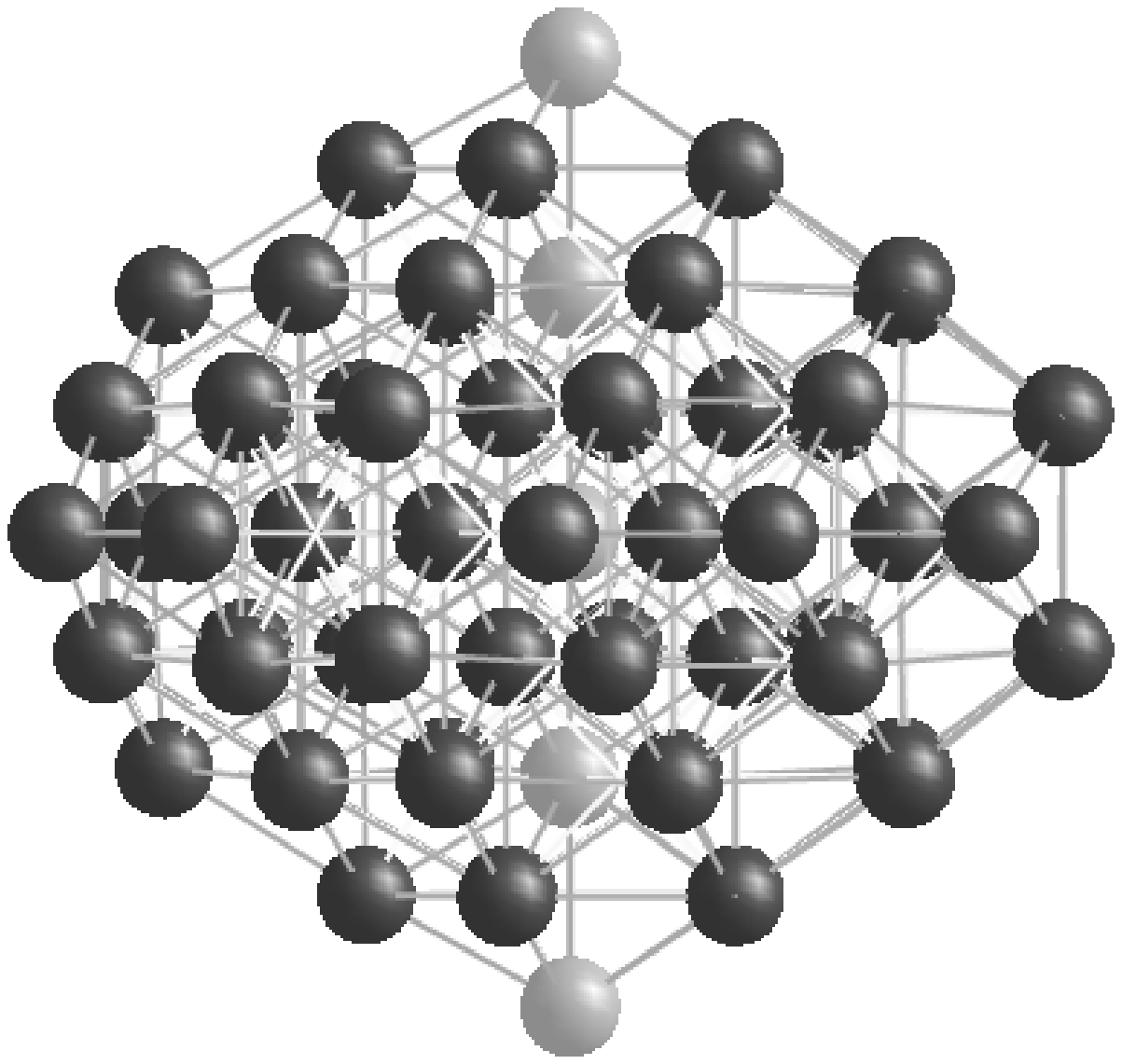}}
\put(5.12,5.3){\includegraphics[width=2.54cm]{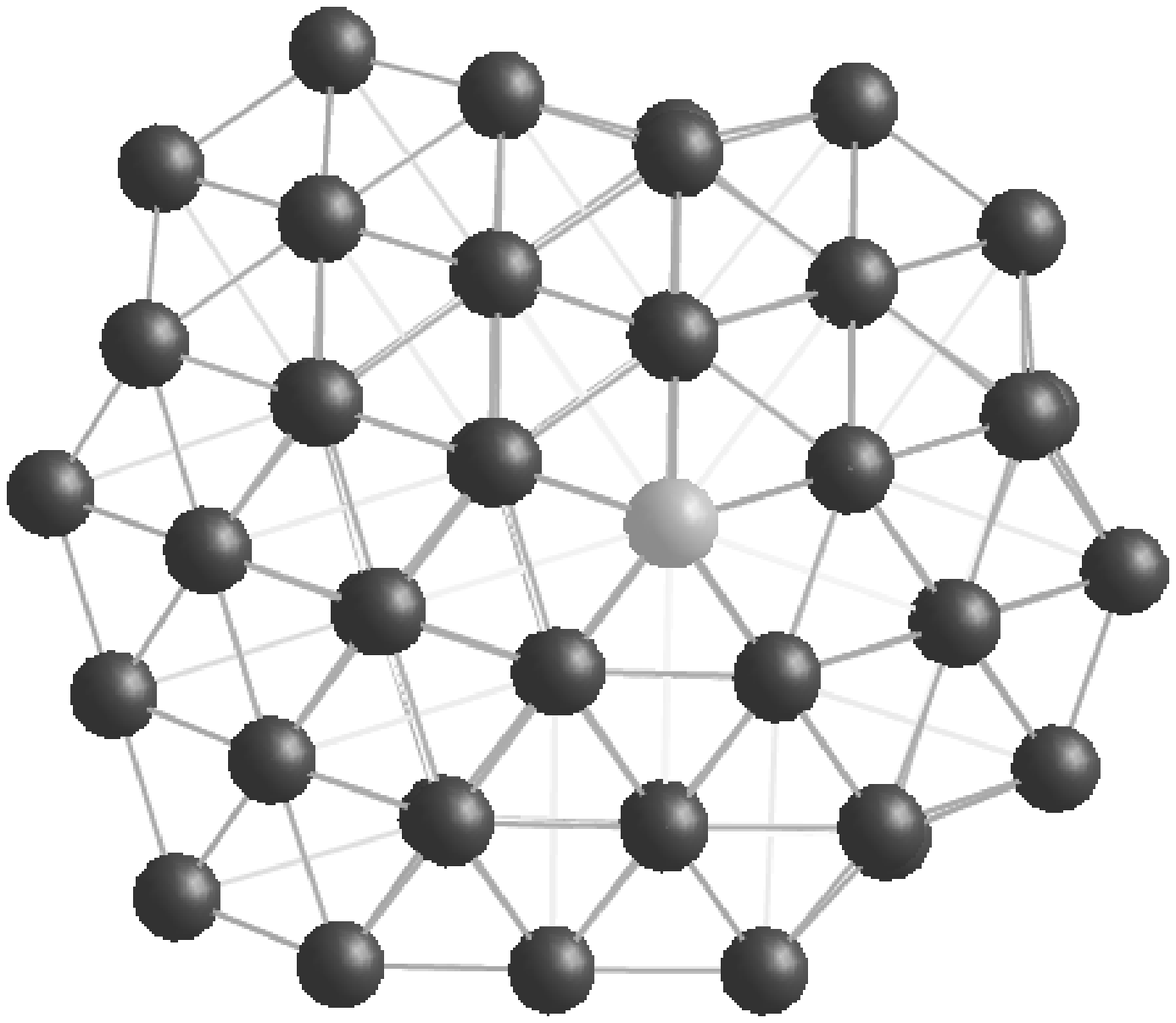} }  

\put(0.0,2.6){\includegraphics[width=2.54cm]{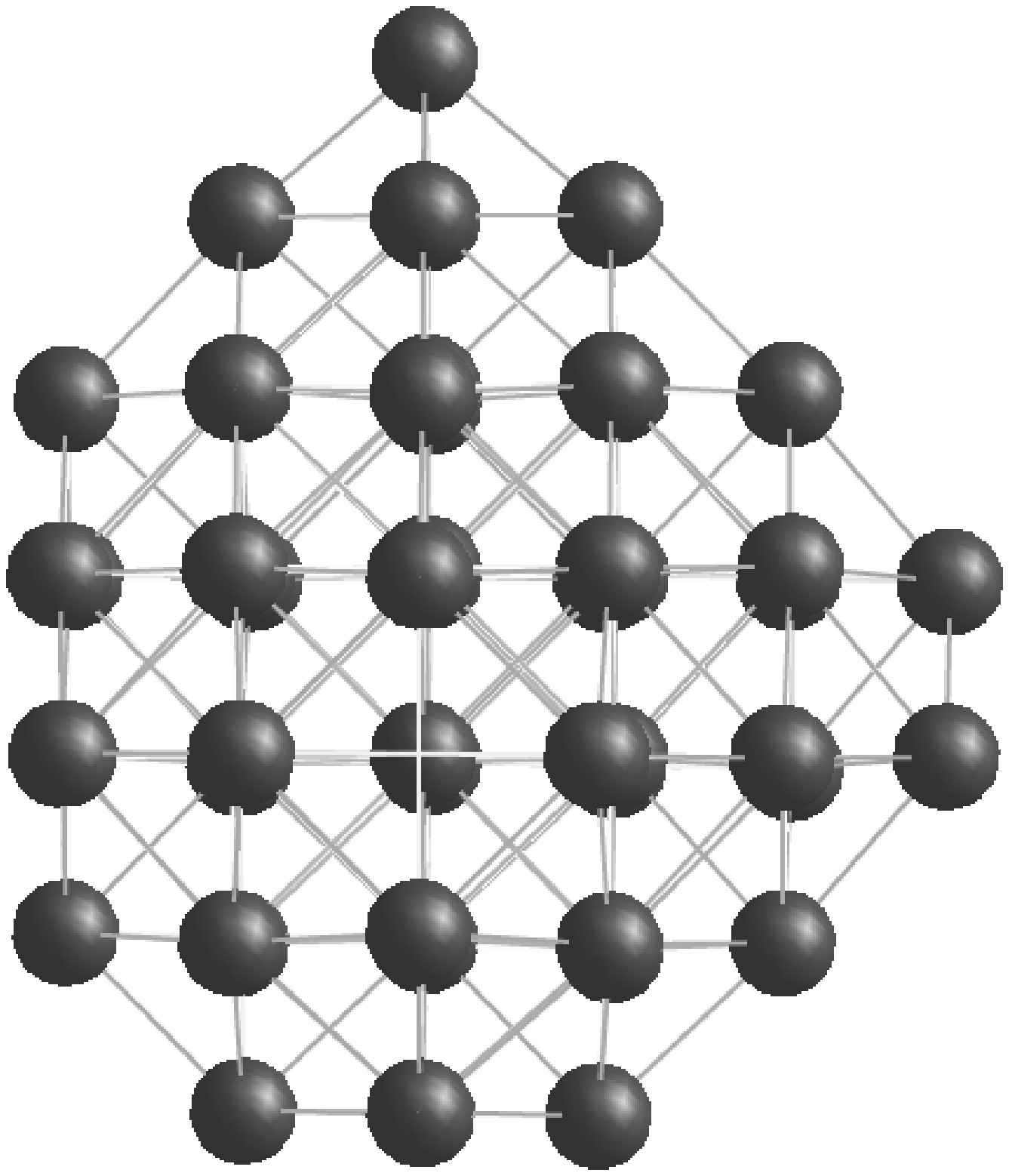}}
\put(2.58,2.6){\includegraphics[width=2.54cm]{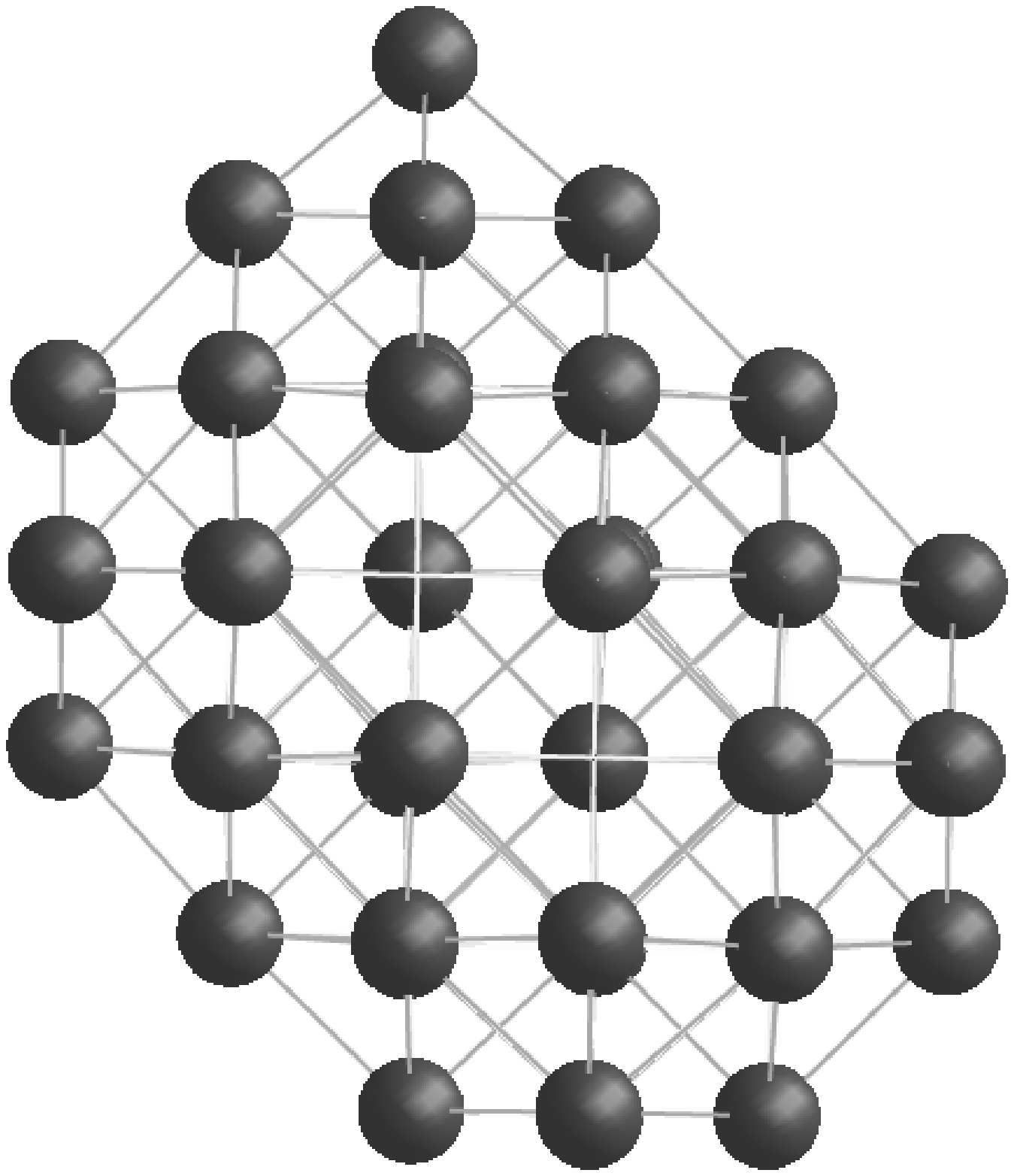}}
\put(5.12,2.6){\includegraphics[width=2.54cm]{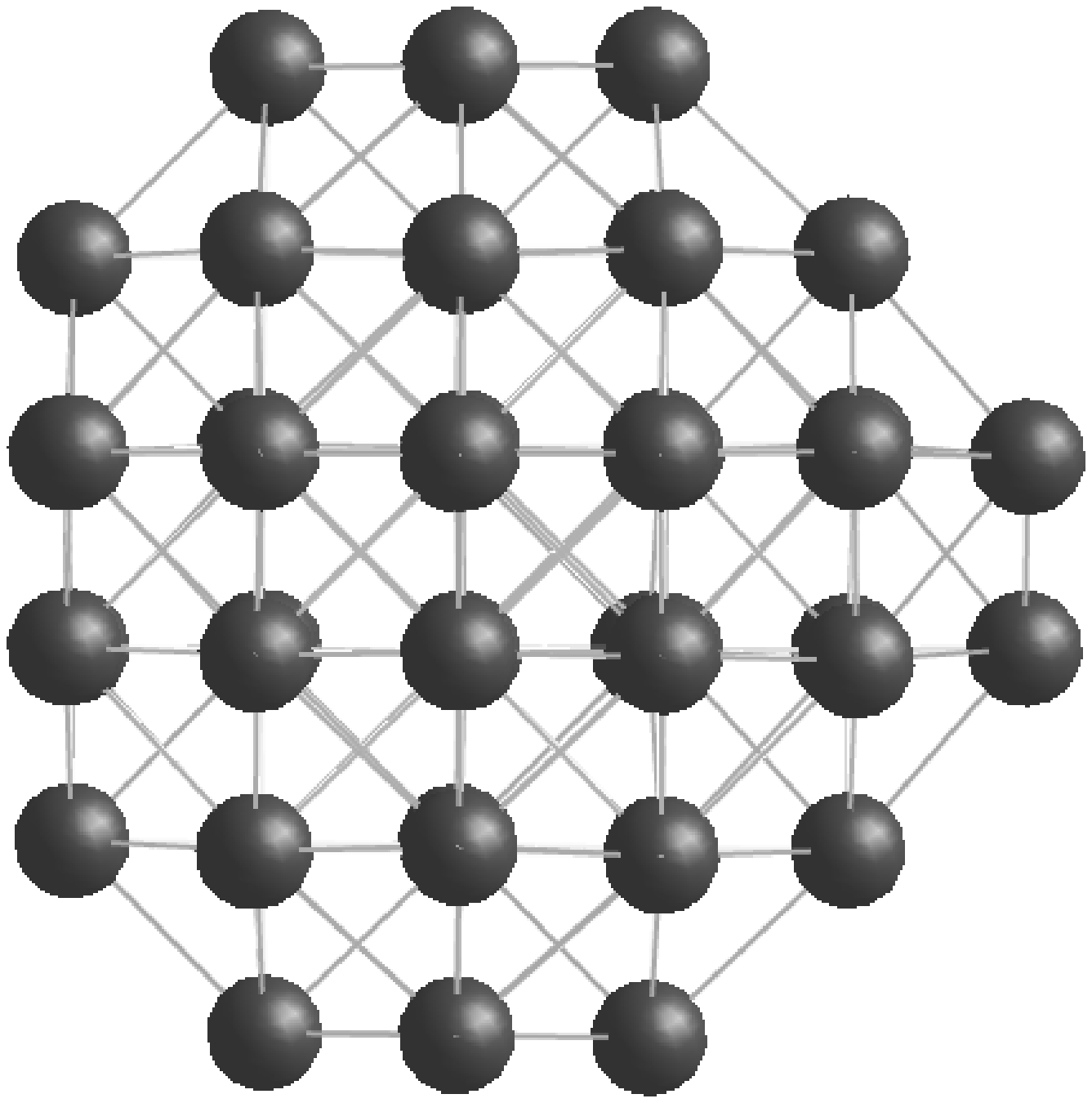} }

\put(0.15,-0.1){\includegraphics[width=2.54cm]{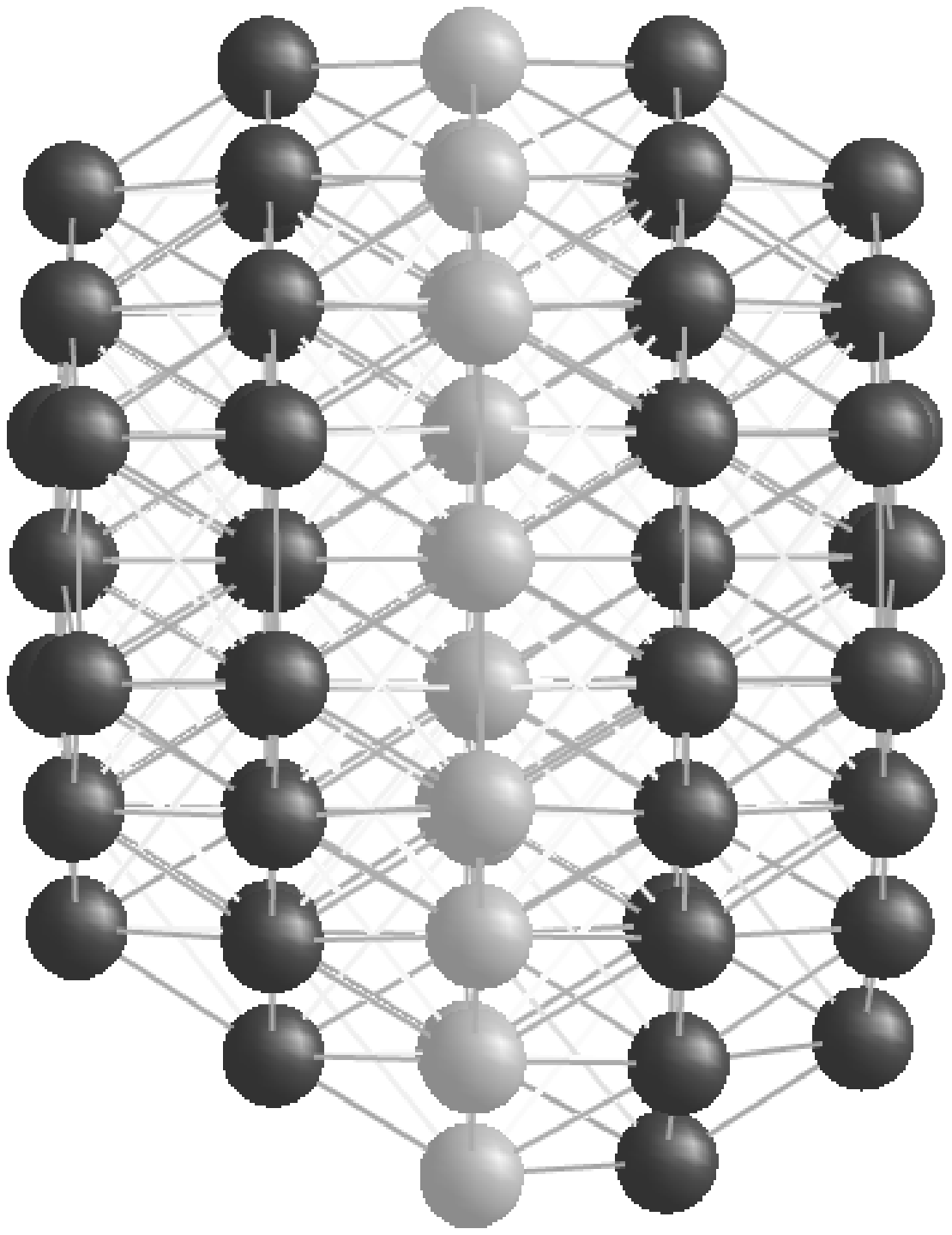}}
\put(2.60,-0.1){\includegraphics[width=2.54cm]{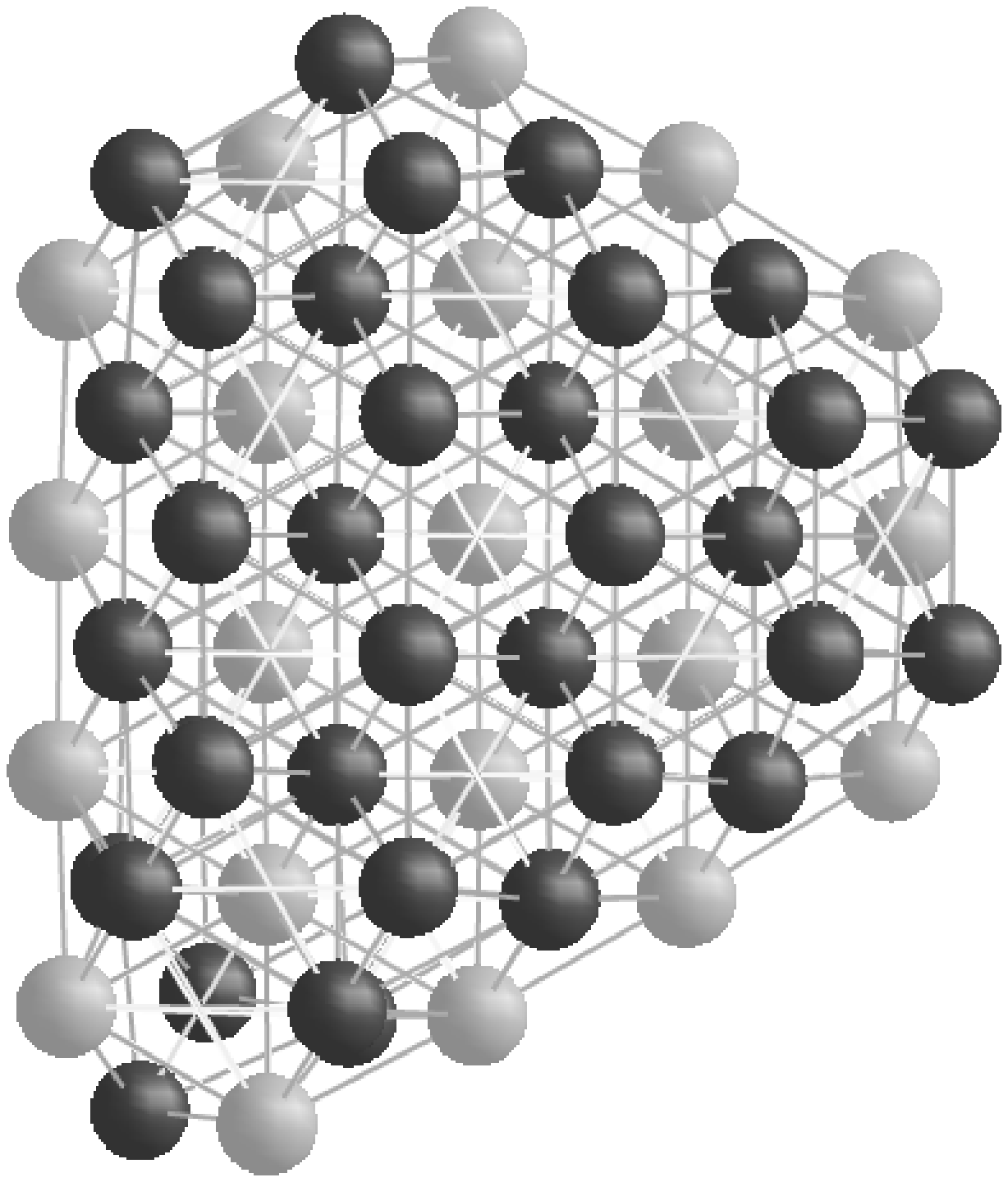}}
\put(5.12,-0.1){\includegraphics[width=2.54cm]{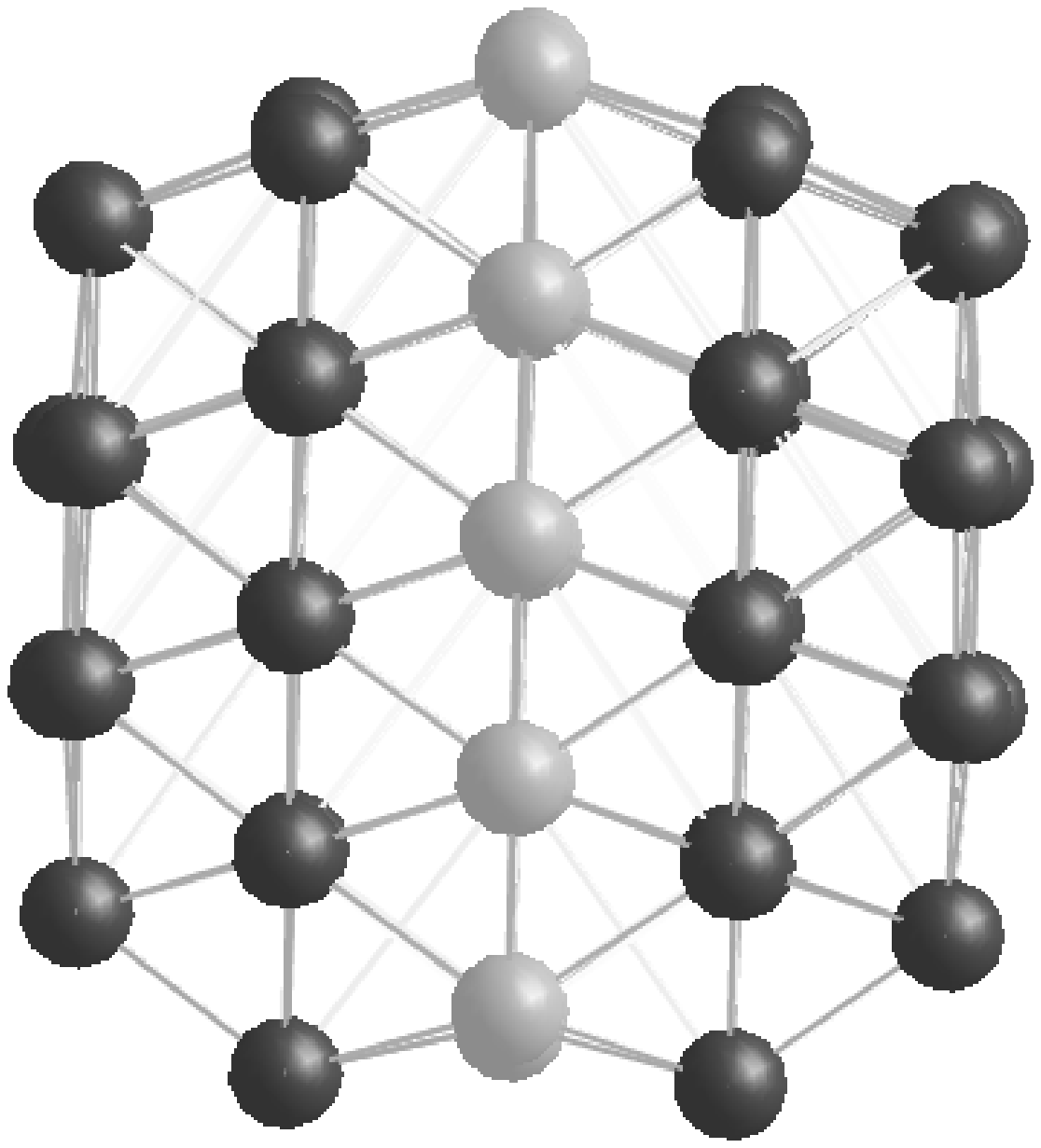} }

\put(3,10.4){amorphous $Au_{54}$}
\put(2.2,7.6){five fold symmetry $Au_{84}$}
\put(3,5.2){single fcc $Au_{55}$}
\put(3,2.4){twinned fcc $Au_{83}$}

\end{picture}\caption{ \label{structures} Illustration of our classification of
the various cluster structures. One row always shows the different
perspectives (along x,y and z) axis of the same cluster. In the case
of the 5-fold structure the atoms on the 5-fold axis are shown in grey
instead of black. In the case of the t-fcc structure the atoms in
the twin plane are shown in grey.}
\end{center}
\end{figure}

Fig.~\ref{energies} shows the central result of our investigation, namely the
energies of all the ground state configurations of the clusters for which
we did a global geometry optimization together with their structure type.
We distinguish the following structure types illustrated in Fig.~\ref{structures}
\begin{itemize}
\item Amorphous: No detectable structure
\item Five fold symmetry (5-fold): A significant subset of atoms has one 5-fold symmetry
      axis.
      Marks Decahedra fall into this class as the special case where the subset comprises
      all the atoms of the cluster.  The large majority of our structures are imperfect in the
      sense that the whole cluster is not invariant under rotations around the
      5-fold axis.
\item Single fcc (s-fcc): the cluster can be cut out of a fcc crystal.
      Perfect octahedra and truncated octahedra fall into this class as a special case.
\item twinned fcc (t-fcc): The cluster consists out of two fcc pieces 'glued' together.
      Twinned octahedra where the two parts are joined together after a rotation
      have been introduced by Raoult~\cite{raoult} and later examined by
      Cleveland~\cite{cleveland} who came  to the conclusion that they are energetically
      not particularly favorable. In our case the structures
      are in most cases not octahedrons but more irregular structures, that just contain
      a microtwin.  Structures of this type have according to our knowledge not been
      reported in the literature up to now.
\end{itemize}

The excess energy to form a cluster out of atoms in the perfect crystal
relative to the number of surface atoms is denoted by $\Delta$ and defined as~\cite{review}
\begin{equation} \label{delta}
\Delta = \frac{E(N)- N \epsilon_{coh}}{N^{2/3}}
\end{equation}
$E(N)$ is the energy of the cluster of $N$ atoms and $\epsilon_{coh}$ the cohesive
energy per bulk Au atom. All energies are calculated with the RGL potential.

Perfect icosahedra (Ih), Marks-decahedra (M-Dh) and truncated
octahedra (TOh) exist only for certain magic numbers. Our global
optimization results of Table~\ref{perfect} show that even in the
case where a gold cluster has a magic number of atoms, the global
minimum is frequently not this perfect structure. In the case where
the global minimum is one of the perfect structures listed in
Table~\ref{perfect} it is not or not much lower in energy than less
symmetric neighboring structures (Fig.~\ref{energies}).

\begin{table}
\begin{tabular}{|  c  |  c | c |  c |} \hline
Number  & magic number & global minimum & relative  \\
 of atoms & type & type &  energy (eV) \\
\hline 13  & Ih (1 shell)& identical& 0. \\
\hline 38  & TOh (4,1) & 5-fold& 0.04 \\
\hline 49  & M-Dh (2,1,2) &identical& 0. \\
\hline 55  & Ih (2 shells) &s-fcc& 0.68 \\
\hline 55  & TOh (5,2) & s-fcc& 1.22 \\
\hline 75  & M-Dh (2,2,2) &identical& 0.\\
\hline 79  & TOh (5,1) & identical&0. \\
\hline 101 & M-Dh (2,3,2) & identical& 0.\\
\hline 116 & TOh (6,2) & s-fcc& 0.41 \\
\hline 140 & TOh (6,1) &identical& 0. \\
\hline 146 & M-Dh (3,2,2) &identical& 0.\\
\hline 147 & Ih (3 shells) &5-fold &1.85\\
\hline 147 & TOh (7,3) &5-fold &2.49 \\
\hline 192 & M-Dh (3,3,2) &identical& 0.\\
\hline 201 & TOh (7,2) &5-fold& 0.007 \\
\hline 309 & Ih (4 shells) &5-fold& 3.64\\
\hline 314 & TOh (8,2) & identical& 0. \\
\hline
\end{tabular} \\
\caption{ \label{perfect} Energies of perfect magic number structures relative to the energy
of the global minimum of the cluster with the same number of atoms found by global
optimization. A energy of 0. indicates that the perfect structure is the global minimum.
The indexing of the M-Dh is the one from ref~\cite{landman}. }
\end{table}

\begin{figure}[h]             
\begin{center}
\setlength{\unitlength}{1cm}
\begin{picture}( 11.,3.)           
\put(0.,0.1){\includegraphics[width=2.7cm]{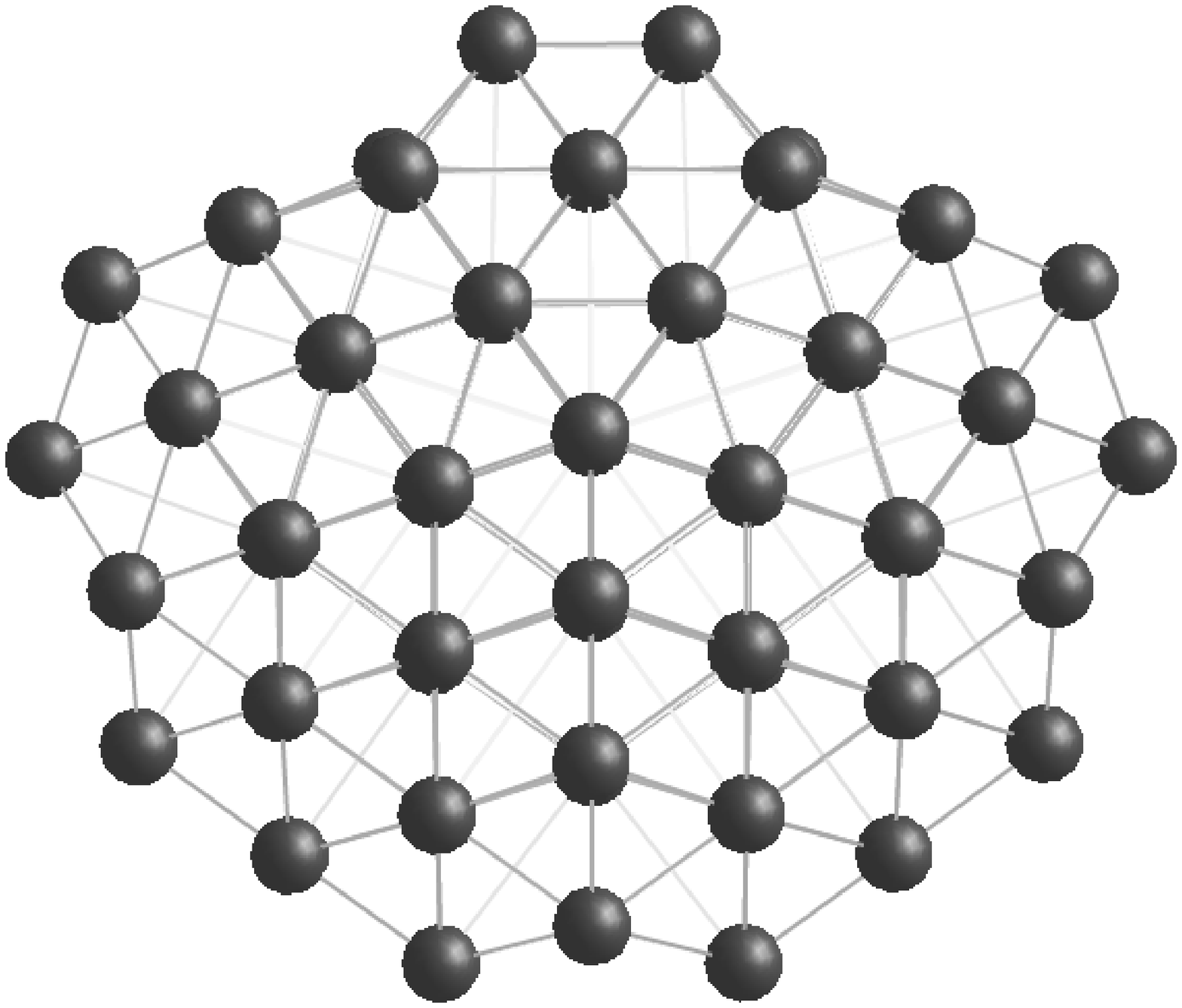}}
\put(3.,0.1){\includegraphics[width=2.7cm]{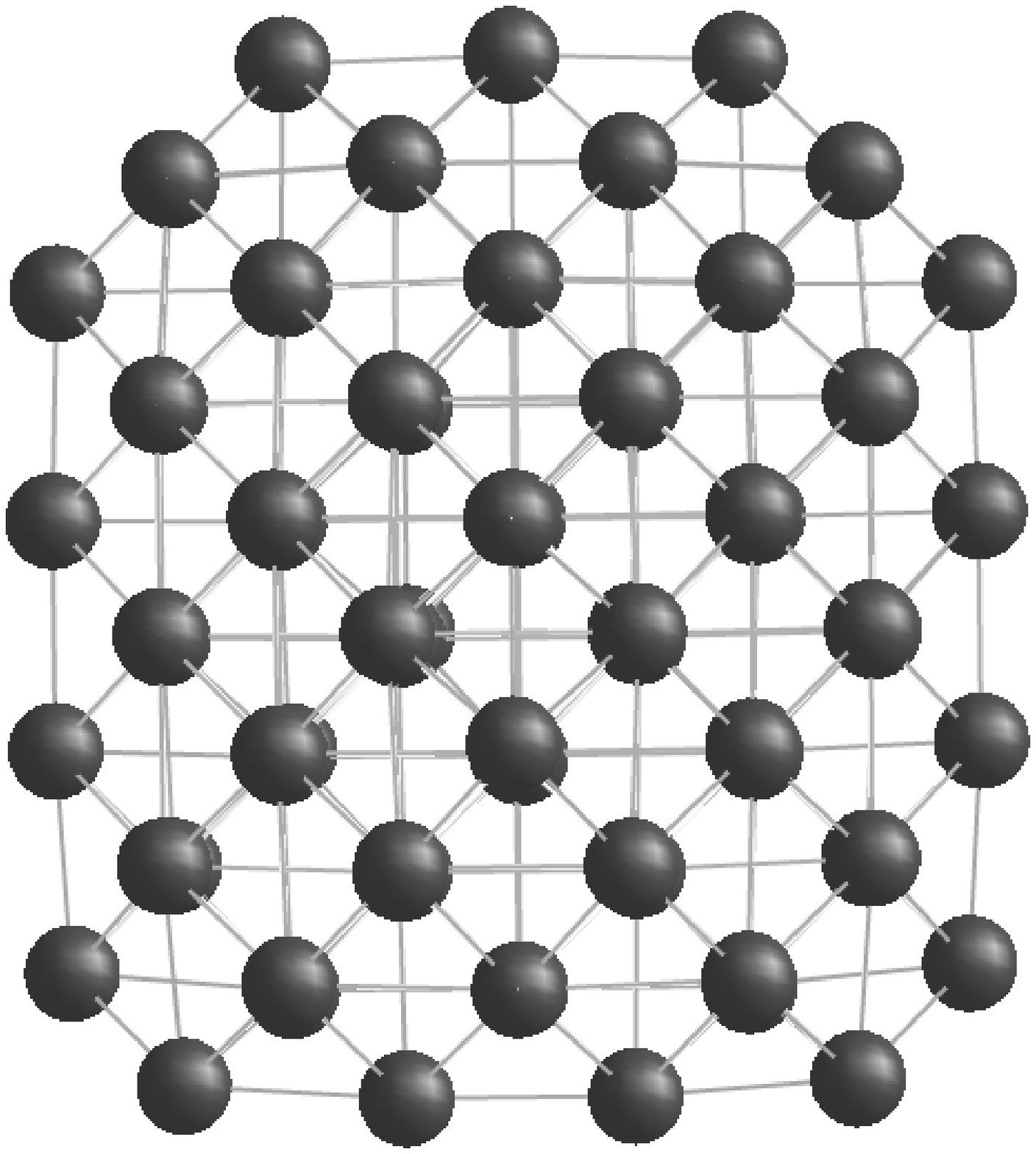}}
\put(5.8,0.1){\includegraphics[width=2.7cm]{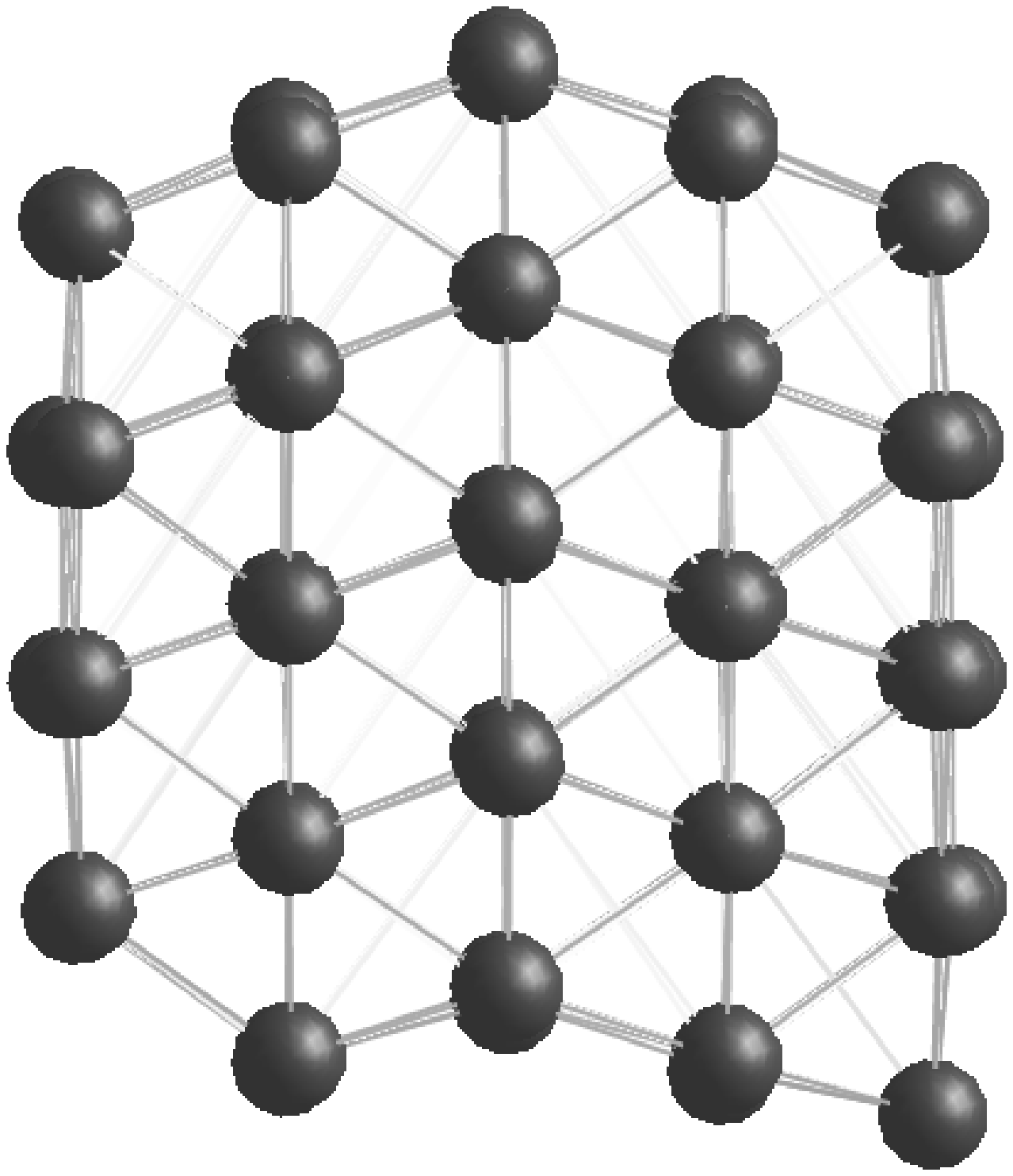} }  %
\end{picture}
\caption{ \label{evolution} The global minimum structure of
$Au_{91}$ (5-fold), $Au_{92}$ (s-fcc) and $Au_{93}$ (t-fcc) }
\end{center}
\end{figure}

Our results show that in most cases  the structure of clusters that
differ just by a single atom is completely different. Thus the rule
that every atom counts is not only valid as hitherto believed for
small clusters of a few dozen atoms that have amorphous structure
but even for clusters of a few hundred atoms. Fig.~\ref{evolution}
shows an example where the structure changes from 5-fold to s-fcc
and t-fcc by the consecutive addition of atoms. Even in cases where
the basic structure type (5-fold, s-fcc, t-fcc) is not modified by
the addition of a single atom on its surface, whole regions change
their structure and the shape of a cluster changes in general.

\begin{figure}[h]             
\begin{center}
\setlength{\unitlength}{1cm}
\begin{picture}( 11.,3.5)           
\put(0.5,0.1){\includegraphics[width=3.cm]{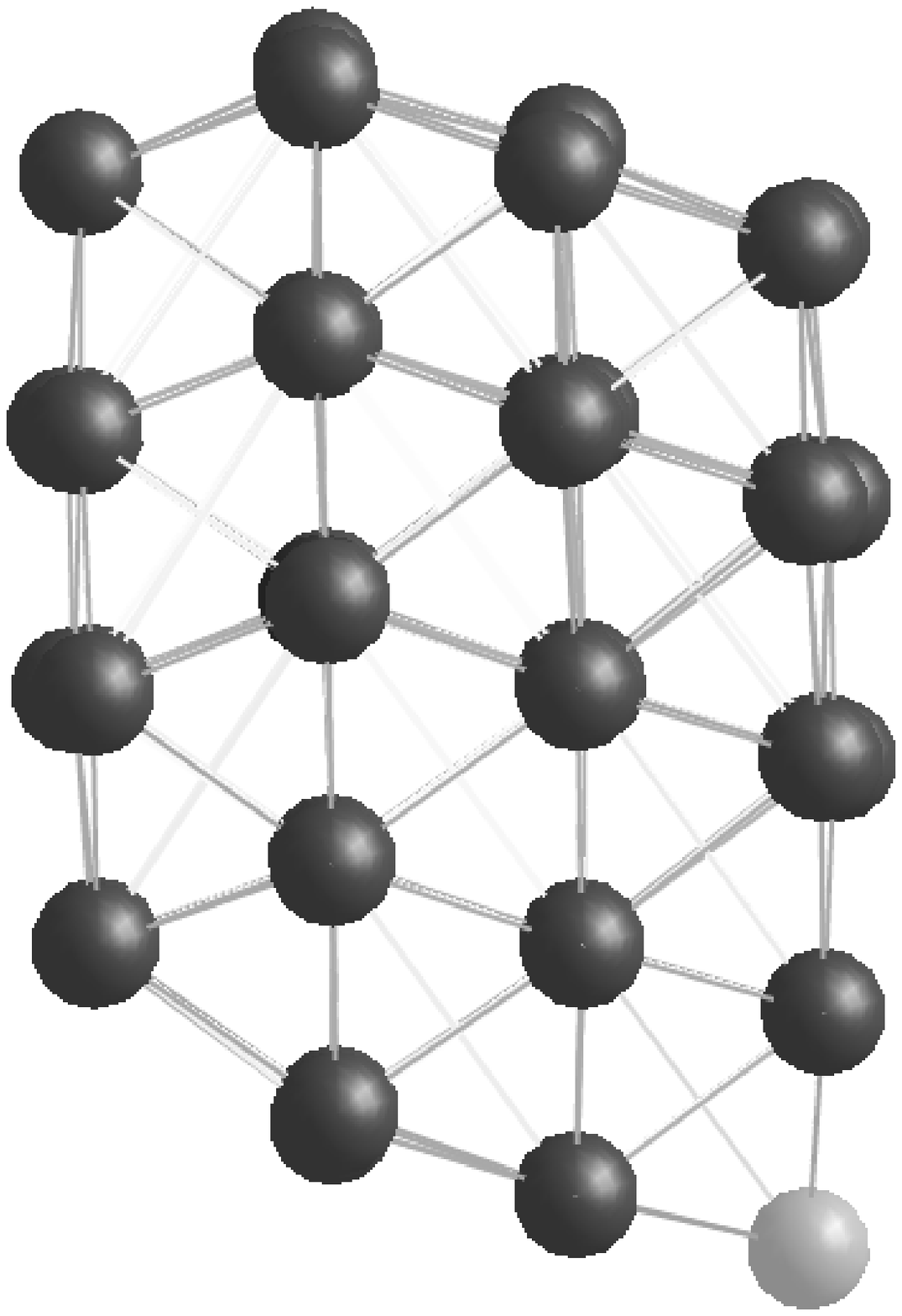}}
\put(4.4,0.1){\includegraphics[width=3.cm]{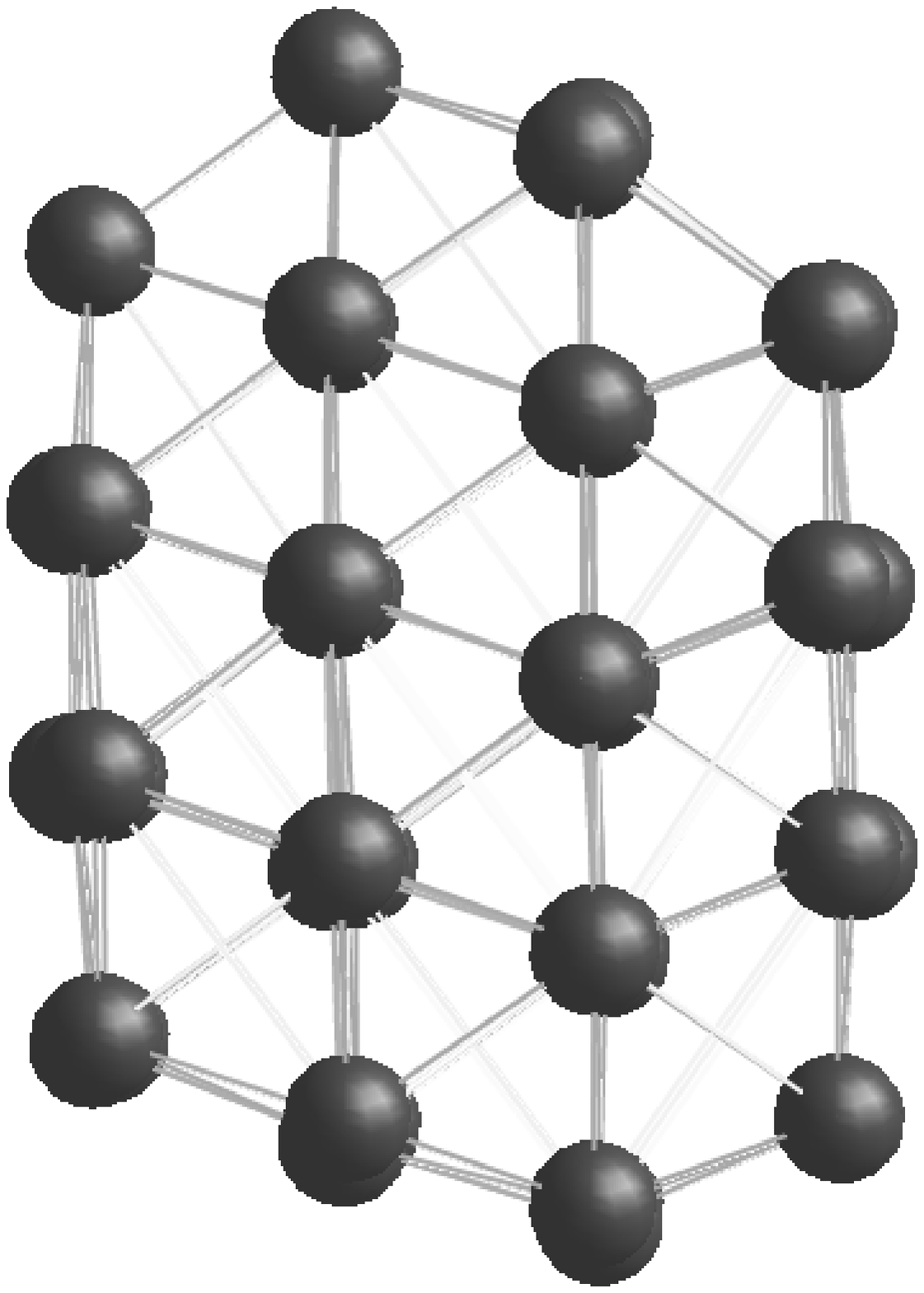}}  
\end{picture}
\caption{ \label{series} The global minima structure of Au$_{62}$
is shown on the left hand side. Taking away a weakly bound atom (shown in grey), we get the
global minima of Au$_{61}$ shown on the right hand side. }
\end{center}
\end{figure}

If clusters that differ only by one atom would in general have
similar structures one could in most cases obtain the global minimum
structure of the $N-1$ atom structure by taking away  a weakly bound
atom of the $N$ atom global minimum structure. We have done this
test and could obtain the $N-1$ ground state only in 48 cases out of
203. One such case is shown in Fig~\ref{series}. 

\begin{figure}[h]             
\begin{center}
\setlength{\unitlength}{1cm}
\begin{picture}( 11.,6.)           
\hspace{-0.2cm} \includegraphics[width=8.5cm]{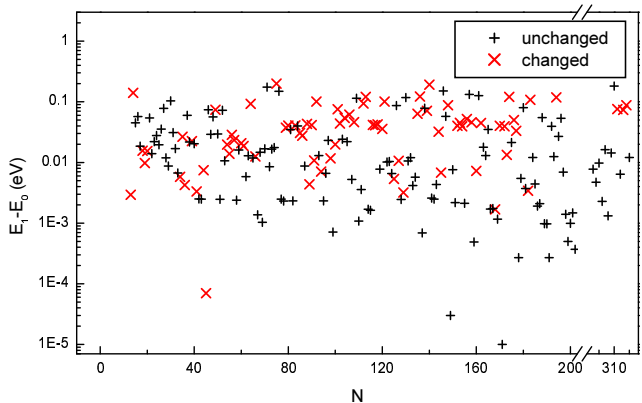}   
\end{picture}
\caption{ \label{secondlowest} The energy difference between the
global minimum and the second lowest local minimum for all the
clusters studied. If both minima share the same structure,
this is indicated by a cross symbol (+);
otherwise by a times symbol ($\times$).
For Au$_{55}$ we have for instance an amorphous structure, previously proposed to
be the ground state~\cite{garzon}, which is only .014 eV higher in energy than
our ground state.}
\end{center}
\end{figure}

It is well known that it is difficult to establish a relation between theoretical
ground state structures and experimental results since in many cases the structure
of clusters is determined by the kinetics of the growing process rather than by
the energetics.  Even if one assumes that thermodynamical equilibrium has been obtained in
an experiment, relating experimental results to simulation results is not
straightforward.
The size of the clusters prepared for diffraction experiments can neither be controlled
up to a single atom, nor can the size be measured after the growth of the cluster
with single atom accuracy in most cases.
Therefore one has experimentally distributions of cluster sizes around
a certain number of atoms.
An additional factor that has to be considered is the fact that the
energy separation between the geometric ground state and the second
lowest local minimum structure is very small.
Fig.~\ref{secondlowest} shows this energy difference for all our
structures. It is of the order of the ambient thermal energy ($k_B T$ = 0.025 eV)
and therefore at room temperature not only the global minimum will
be encountered but also a few other low energy configurations.
Similar small energy differences were also found for a few selected small
gold clusters by Soler et al.~\cite{soler} and for silicon clusters by Hellmann
et al~\cite{hellmann2} using rather accurate density functional and
Quantum Monte Carlo methods. The
existence of a large number of structures that are energetically
very close to the global minimum is also illustrated by the fact
that within an energy interval of .1 eV above the global minimum
there are on average 12 configurations for the clusters studied in this work.
All these facts show that in experiment one can not expect to observe a
single structure but rather a distribution of structures as it is indeed the
case~\cite{koga,li}.

Like for silicon clusters~\cite{hellmann2} it is expected that
entropy effects can change the energetic ordering. i.e. the order
given by the free energy at room temperature or somewhat above is different from the
energetic ordering at zero temperature. Highly symmetric
configurations become disfavored with respect to less symmetric
configurations at finite temperatures. We  found many low symmetry
configurations that are very close in energy to high symmetry global
minimum configurations. For the M-Dh $Au_{192}$ cluster
(Table\ref{perfect}) there is for instance another configuration
that is only 0.04 eV higher in energy and which has only one mirror
plane. Even though this small energy difference is certainly not
reliable given by the RGL potential it is to be expected that DFT or
even Quantum Monte Carlo calculations would give low symmetry
configurations that are very close in energy to certain high symmetry
configurations.

Given the fact that the error bars of the RGL potential are larger
than the energy differences between the global minimum structure and
other low energy structures we can certainly not claim that the
global minima structures that we found are the true global minima
structures. What is however true and does not depend on the form of
the interatomic potential is that there exist completely different
types of structures that are very close in energy.  This is due to the fact
that a cluster can not satisfy all the conditions that would lead to
a lowering of the energy, namely a small surface area, mainly (111) surfaces,
few grain boundaries and little strain in a fcc like core region.
Favoring one condition at the expense of another gives similar energies.

Taking out an atom is a significant perturbation which
induces in many cases a complete structural change of the entire
cluster. This is due to the fact that the change in the binding energy
$E(N)- N \epsilon_{coh} $ i.e. $E(N+1)-E(N)-\epsilon_{coh}$ is the
order of 0.15 eV for the cluster sizes we studied, which is larger
than the typical energy difference between the global minimum and
the second-lowest local minimum( see Fig.~\ref{secondlowest}).

The fact that taking out a single or a few atoms in our global geometry
optimization induces a
structural change can also be understood by a geometric argument.
All our low energy structures have the property that they have
rather smooth surfaces. Taking out one or several atoms will at some
point lead to non-smooth surfaces that have steps or holes and thus
to cluster shapes that are energetically unfavorable.

The structures of the ground state and the first excited configuraration
are available as supplementary material. We thank the Swiss National
Science Foundation and the CSC (China Scholarship Council)
for the financial support of our research work and the CSCS for computing time.


\begin{thebibliography}{99}
\bibitem{landman} C. Cleveland and U. Landman,
                     J. Chem. Phys. {\bf 94 }, 7376 (1991).
\bibitem{zeng} S. Yoo and X.C. Zeng, Angew. Chem. Int. Ed. {\bf 44}, 1491
(2005).
\bibitem{hellmann1} S. Goedecker, W. Hellmann and T. Lenosky,
           Phys. Rev. Lett. {\bf 95}, 055501 (2005).

\bibitem{Au55}  H.-G. Boyen, G. K\"{a}stle, F. Weigl, B. Koslowski, C. Dietrich,
P. Ziemann, J. P. Spatz, S. Riethm\"{u}ller, C. Hartmann, M.
M\"{o}ller, G. Schmid, M. G. Garnier and P. Oelhafen, Science, {\bf
297}, 1533 (2002).

\bibitem{garzon} K. Michaelina, N. Rendon and I. Garzon,
                Phys. Rev. B {\bf 60}, 2000 (1999).
\bibitem{doye}  J. Doye and D. Wales,
                  New J. Chem.  773 (1998).
\bibitem{rgl}  V. Rosato, M. Guillope and B. Legrand,
                    Phil. Mag. A {\bf 59 }, 321 (1989).

\bibitem{raoult}  B. Raoult, J. Farges. M. de Feraudy and G. Torchet,
                    Phil. Mag. B {\bf 60 }, 881 (1989).

\bibitem{baletto} F. Baletto, R. Ferrando, A. Fortunelli, F. Montalenti and C. Mottet, J. Chem. Phys. {\bf 116}, 3856
(2002).

\bibitem{carvalho} F. Negreiros, E. Soares and V. Carvalho, Phys. Rev. B {\bf 76}, 105429-1
(2007).

\bibitem{minhop} S.Goedecker, J. Chem. Phys. {\bf 120}, 9911 (2004).

\bibitem{roy} S. Roy, S. Goedecker, and V. Hellmann
Phys. Rev. E {\bf 77}, 056707 (2008).
\bibitem{bigdft} L. Genovese et al., J. Chem. Phys., to be published.

\bibitem{cleveland} C. Cleveland, U. Landman, T. Schaff and M. Shafigullin,
           Phys. Rev. Lett. {\bf 79}, 1873 (1997).

\bibitem{review} F. Baletto and R. Ferrando, Rev. Mod. Phys.  {\bf 77}, 371
(2005).

\bibitem{soler}  J. M. Soler, M. R. Beltran, K. Michaelian, I. L. Garzon, P. Ordejon, D. Sanchez-Portal and E. Artacho,
                  Phys. Rev. B {\bf 61} 5771 (2000).
\bibitem{hellmann2}
W. Hellmann, R. G. Hennig, S. Goedecker, C. J. Umrigar, Bernard
Delley and T. Lenosky, Phys. Rev. B {\bf 75}, 085411 (2007).

\bibitem{koga} K. Koga, T. Ikeshoji and K. Sugawara, Phys. Rev. Lett.  {\bf 92}, 115507-1
(2004).

\bibitem{li} Z. Y. Li, N. P. Young, M. Di Vece, S. Palomba, R. E. Palmer, A. L. Bleloch, B. C. Curley, R. L. Johnston, J. Jiang and J. Yuan, Nature {\bf
451}, 46 (2007).


\end{thebibliography}
\end{document}